\documentclass[prl,twocolumn,showpacs]{revtex4}
\usepackage{amsmath}
\usepackage{amssymb}
\usepackage{graphicx}

\begin{document}
\title{Tunable Whispering Gallery Mode Resonators for Cavity
        Quantum Electrodynamics}
    \author{Y.~Louyer}
    \author{D.~Meschede}
    \author{A.~Rauschenbeutel}
    \email{rauschenbeutel@iap.uni-bonn.de}
    \affiliation{Institut f\"ur Angewandte Physik, Universit\"at
Bonn,
    Wegelerstr.~8, D-53115 Bonn, Germany}
\date{\today}
\pacs{42.55.Sa, 03.65.Ge, 42.50.Pq}

\begin{abstract}
We theoretically study the properties of highly prolate shaped
dielectric microresonators. Such resonators sustain whispering
gallery modes that exhibit two spatially well separated regions
with enhanced field strength. The field per photon on the
resonator surface is significantly higher than e.g.~for equatorial
whispering gallery modes in microsphere resonators with a
comparable mode volume. At the same time, the frequency spacing of
these modes is much more favorable, so that a tuning range of
several free spectral ranges should be attainable. We discuss the
possible application of such resonators for cavity quantum
electrodynamics experiments with neutral atoms and reveal distinct
advantages with respect to existing concepts.
\end{abstract}

\maketitle

Optical microresonators have attracted much interest in the field
of linear and nonlinear optics as well as in cavity quantum
electrodynamics, as summarized in the recent review by K.~Vahala
\cite{Vahala}. One of the reasons for this considerable attention
is the prime importance of a small mode volume $V$ for such
applications. So far, optical microresonators based on Fabry-Perot
cavities, photonic crystals, and whispering gallery modes have
been realized \cite{Vahala}. In addition to $V$, certain
applications, like e.g.~cavity quantum electrodynamics
\cite{Berman}, require also a minimization of the ratio $V/Q$,
where the quality factor $Q$ is determined by the energy storage
time in units of the optical period.

Very high $Q$-values have been realized for Fabry-Perot cavities
with ultrahigh reflectance mirrors \cite{RempeLalezari}. However,
the mirror components are involved and costly to manufacture.
Furthermore, due to the modular construction, such resonators
require an elaborate active stabilization of their resonance
frequency. Photonic crystal based resonators have a much better
passive stability and yield extremely small mode volumes. Even
though the experimentally realized $Q$-values fall well below the
theoretical limits, they therefore reach record values for $V/Q$
\cite{Akahane}. However, the problem of tuning the resonance
frequency of such structures has not been solved to this date.

The highest $Q$-values for optical resonators to date have been
realized with whispering gallery modes (WGMs) in fused silica
microspheres \cite{Gorodetsky}, where continuous total internal
reflection confines the light to a thin equatorial ring near the
surface of the sphere. In combination with their relatively small
mode volume of $V\approx 1000\, \lambda^3$, where $\lambda$ is the
optical wavelength, these modes are therefore ideal candidates for
cavity quantum electrodynamics experiments.

Optical probing and output coupling requires phase matching of the
WGMs to propagating light fields. This has been achieved with
various methods, all of which use an auxiliary dielectric
structures (prisms \cite{Ilchenko}, eroded or tapered optical
fibers \cite{SerpenguzelDubreuil,Knight}, or fiber/prism hybrid
systems \cite{Yao}) that are introduced into the external
evanescent field of the WGM in the vicinity ($\lesssim \lambda$)
of the sphere surface. Using such a set-up, strong coupling
between atoms in a dilute vapor and a microsphere WGM has been
realized \cite{Vernooy}. However, in spite of considerable
research activities, a {\em controlled} strong coupling of a
suitable dipole emitter to a microsphere WGM has not yet been
accomplished. The main difficulty to be overcome consists in
controllably placing the emitter as close as possible to the
dielectric surface. In particular, due to the two-dimensional
character of equatorial WGMs, the prism or fiber coupler severely
limits the mechanical and optical access.

Another practical difficulty stems from the limited tunability of
microsphere WGMs. For a typical microsphere of $\varnothing\approx
50\ \mu$m, tuning over one free spectral range (FSR) requires a
change of the optical path length of the WGM of about $4\times
10^{-3}$. On the other hand, the temperature dependence of the
refractive index of silica is only $\partial n/\partial T\approx
1.3\times 10^{-5}\ \mathrm{K}^{-1}$. This index change is the
dominant temperature effect in silica. Therefore, a temperature
variation can only be used for fine tuning in this case. The only
practicable solution for tuning is thus to elastically deform the
sphere through mechanical strain. For a $\varnothing \approx 80\
\mu$m microsphere, tuning over one half of an FSR has been
achieved with this method, limited by the mechanical damage
threshold of the resonator set-up \cite{Klitzing}. In addition,
for a $\varnothing\approx 50\ \mu$m microsphere with $Q\approx
10^9$ the resonance linewidth is only $10^{-7}$ of an FSR.
Therefore, even if such a tuning device reached a full FSR tuning
range, it would be a challenging task to guarantee the
corresponding passive stability.

\begin{figure}[t]
          \centering
          \includegraphics [scale=0.25]{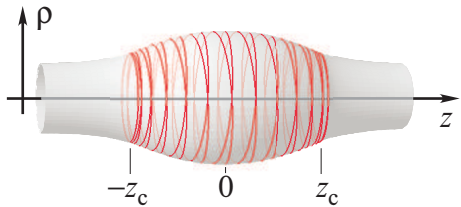}
          \caption{Sketch of the resonator geometry. The ray path
          corresponding to the whispering gallery modes under
          consideration is schematically indicated. Note that the
          thickness variation of the resonator structure along
          the $z$-axis is amplified for illustration purposes.}
          \label{fig:Resonator}
\end{figure}

Here, we propose the use of WGMs in highly prolate dielectric
resonators with a cylindrical symmetry. Of course, such a
structure also sustains ``equatorial'' WGMs, i.e.~with a
corresponding ray path that is closed after one revolution around
the resonator axis. In the following, however, we will consider
modes for which the light spirals back and forth along the
resonator axis between two turning points, separated by
$2z_\mathrm{c}$, see fig.~\ref{fig:Resonator}.

From this very simple ray path picture it is already apparent that
such modes exhibit two spatially separated caustics located at
$\pm z_\mathrm{c}$ with an enhanced field strength. Note that, in
analogy to a charged particle in a magnetic bottle, the light is
in fact confined along the $z$-axis between the two turning points
by an angular momentum barrier. For this reason, we have adopted
the denomination ``bottle modes'' from \cite{Summetsky}, where
such a resonator shape was theoretically constructed from an
equidistant spectrum of eigenmodes using WKB approximation.

We will now present a full scale wave equation calculation for
such a resonator shape and theoretically explore important
resonator properties including field distribution, mode volume,
and tunability. We consider a resonator profile which is
approximately parabolic along $z$:
\begin{equation}\label{eq:profile}
R(z)\approx R_0\left(1-\frac{1}{2}(\Delta k\, z)^2\right)\ ,
\end{equation}
where we take $R_0\approx 8\ \mu$m as the maximal radius and
$\Delta k^2\approx 10^{-5}\ \mu\mathrm{m}^{-2}$ for the curvature
of the profile. We use the method of adiabatic invariants
\cite{Percival}, well known from the description of the dynamics
in magnetic bottles. Indeed, due to the small variation of the
resonator radius (amplified in Fig.~\ref{fig:Resonator}), $dR/dz
\ll 1$, and since we are considering maximum angular momentum
modes located close to the surface of the resonator, the radial
component $k_{\rho}$ of the wave vector is negligible with respect
to $k_{z}$ and $k_{\varphi}$. Therefore, the total wave number is
\begin{equation}\label{eq:paraxial_approx}
k=(k^2_{z}+k^2_{\varphi})^{1/2}=2\pi n/\lambda\ ,
\end{equation}
where $n$ and $\lambda$ are the effective index of refraction and
the wavelength in vacuum, respectively. Now, due to cylindrical
symmetry, the projection of the angular momentum onto the $z$-axis
is a conserved quantity, $\partial_z\, k_{\varphi}(z)R(z)=0$.
Furthermore, the axial component of the wave vector vanishes in
the caustics, $k_{z}(\pm z_\mathrm{c})=0$, so that $k_\varphi(\pm
z_\mathrm{c})=k$. The azimuthal and axial components of the wave
vector thus read:
\begin{align}\label{eq:phiwavevectorcomponent}
    k_\varphi(z)&=kR_\mathrm{c}/R(z)\quad \mathrm{and}\\
    k_{z}(z) &= \pm k\sqrt{ 1 - \bigl( R_\mathrm{c}/R(z)\bigr)^2 }
    \ ,\label{eq:zwavevectorcomponent}
\end{align}
where $-z_\mathrm{c}\leq z\leq z_\mathrm{c}$ and
$R_\mathrm{c}=R(z_\mathrm{c})$.

In order to check that the paraxial approximation of a negligible
radial wave vector component is well justified for the resonator
geometry of eq.~(\ref{eq:profile}), we use the relation
$k_\rho=(dR/dz)k_z$. For $z_\mathrm{c}\approx 70~\mu$m this yields
$|k_{\rho}(z)|<6\times 10^{-3}|k_{z}(z)|$ in the region
$-z_\mathrm{c}\leq z \leq z_\mathrm{c}$. Inserting this value as a
correction into eq.~(\ref{eq:paraxial_approx}), we estimate the
error in eq.~(\ref{eq:zwavevectorcomponent}) to be smaller than
$2\times 10^{-5}$.

Using the adiabatic approximation, we will now establish the wave
equation and determine the eigenfunctions of the resonator. Due to
the cylindrical symmetry, the azimuthal part of the wave equation
can be separated with solutions proportional to $\exp(im\varphi)$,
where $m$ is the azimuthal quantum number. The solution can thus
be written $\Psi(\rho,z)\exp(im\varphi)$. The radial quantum
number $p$ will be fixed to its minimum value $p=1$, corresponding
to modes located at the surface of the resonator. Now, using the
adiabatic approximation along the $z$-axis, $\Psi(\rho,z)$ can be
separated in a product of two functions $\Phi(\rho,z)Z(z)$, where
$\Phi$ is the solution of a Bessel equation,
\begin{equation}\label{eq:radial_wave_equation1}
\frac{\partial^{2}\Phi}{\partial \rho^{2}} +
\frac{1}{\rho}\frac{\partial \Phi}{\partial \rho} + \left(
k_{\varphi}^{2} - \frac{m^{2}}{\rho^{2}} \right) \Phi = 0\ .
\end{equation}
Here, $k_{\varphi}$ is given by
eq.~(\ref{eq:phiwavevectorcomponent}) and, using
$kR_\mathrm{c}=m$, equals $k_{\varphi}(z)\!=m/R(z)$. Insertion
into the above equation yields
\begin{equation}\label{eq:radial_wave_equation2}
\frac{\partial^{2}\Phi}{\partial \rho^{2}} +
\frac{1}{\rho}\frac{\partial \Phi}{\partial \rho} + m^2\left(
\frac{1}{R(z)^2} - \frac{1}{\rho^{2}} \right) \Phi = 0\ .
\end{equation}
In order to determine the radial wave function from this equation,
we first have to fix the boundary conditions at the resonator
surface, which depend on the polarization of the mode. We consider
a TM~mode, i.e.~linear polarization with the electric field
parallel to the $z$-axis. In this case, the solution of equation
(\ref{eq:radial_wave_equation2}) can be written
\begin{equation}
\Phi_m(\rho,z) =
    \begin{cases}
    A_{m} J_{m}(k_{\varphi}\rho) & ( \rho < R(z) ) \\
    H_{m}^{(2)}(\frac{k_{\varphi}\rho}{n}) + S_{m}
H_{m}^{(1)}(\frac{k_{\varphi}\rho}{n}) & ( \rho> R(z) )
    \end{cases}
    ,
\end{equation}
where $J_{m}$ and $H_{m}^{(1,2)}$ are the Bessel and Hankel
functions, respectively. The coefficients $A_{m}$ and $S_{m}$ are
determined such that they satisfy the matching conditions for
$\Phi$ and its derivative, $\partial \Phi/\partial\rho$.

The differential equation for $Z$ also depends on the resonator
profile $R(z)$. In order to find an analytic solution, we choose
the explicit form $R(z)=R_{0}/\sqrt{ 1 + ( \Delta k z )^{2}}$
which, given that $(\Delta k z)^2<0.05$ in our case, realizes the
parabolic profile of equation (\ref{eq:profile}) to a very good
approximation. Using this form, the problem reduces to a simple
harmonic oscillator,
\begin{equation}\label{eq:axial_wave_equation}
\frac{\partial^{2}Z}{\partial z^{2}} + \left( k^{2} -
\frac{m^{2}}{R_0^{2}} - \frac{m^{2}\Delta k^{2}}{R_0^{2}}z^{2}
\right) Z = 0.
\end{equation}
In analogy to the harmonic oscillator problem, we can therefore
identify the total and the potential energy as $E=k^2-m^2/R_0^{2}$
and $V(z)=(\Delta E_m z/2)^2$, respectively, where we have set
$\Delta E_m = 2m\Delta k/R_{0}$. Furthermore, the condition that
$Z$ be square integrable leads to a discrete set of energy levels
$E_{mq}=(q + 1/2) \Delta E_m$, where $q$ is the axial quantum
number, corresponding to the number of nodes of the wave function
along $z$. This allows us to deduce the allowed eigenvalues
\begin{equation}\label{eq:eigenwavenumber}
k_{mq} = \left[ m^{2}/R_{0}^{2} + \left( q + 1/2 \right)\Delta E_m
\right]^{1/2}\ ,
\end{equation}
with the corresponding solution of equation
(\ref{eq:axial_wave_equation}),
\begin{equation}\label{eq:axial_wave_function}
Z_{mq}(z) = C_{mq} H_{q}\!\left( \sqrt{\frac{\Delta E_m}{2} }\,z
\right) \exp\!\left(- \frac{\Delta E_m}{4}\,z^{2}\right)\ ,
\end{equation}
where $H_{q}$ is the Hermite polynomial of order $q$ with the
normalization constant $C_{mq}=\left[ \Delta E_m / (\pi
2^{2q+1}(q!)^2) \right]^{1/4}$.

\begin{figure}[t]
          \centering
          \includegraphics [scale=0.18]{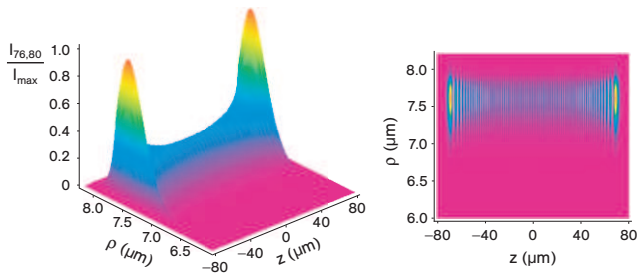}
          \caption{Intensity distribution of the bottle mode.}
          \label{fig:3DIntensity}
\end{figure}
From the wave function $\Psi_{mq}(\rho,z)=\Phi_m(\rho,z)Z_{mq}(z)$
we can now calculate the intensity distribution
$I_{mq}(\rho,z)\propto |\Psi_{mq}(\rho,z)|^2$ of the bottle modes.
We consider the case where the wavelength $\lambda_{mq}$ matches
the D2 line of Cs at 852~nm. We choose the resonator radius at the
caustics to be $R_\mathrm{c}=7.8$~$\mu$m, compatible with a
quality factor of $Q\approx 10^9$ \cite{Buck}. Moreover, we set
$\Delta k=0.0032$~$\mu$m$^{-1}$, $q=80$, and $m=76$, i.e.~the
highest possible $m$ value. Figure~\ref{fig:3DIntensity} shows
$I_{mq}(\rho,z)$ for the corresponding bottle mode. The intensity
in the caustics, located at $z_\mathrm{c}=\pm 69$ $\mu$m, is
enhanced by a factor of four compared to the peak value around
$z=0$.

The intensity distribution directly yields the mode volume $V$ of
the bottle modes. It is obtained by normalizing $I_{mq}(\rho,z)$
to unity and by integrating it according to:
\begin{equation}\label{eq:mode_volume}
V_{mq} = \int\!\!\! \int\!\!\! \int_0^{\rho_0(z)}\!\!\!\!\!
\varepsilon(\rho,z) \frac{I_{mq}(\rho,z)}{I_{mq}^{\mathrm{max}}}\,
\rho\,d\rho\,d\varphi\,dz\ ,
\end{equation}
where $\varepsilon(\rho,z)=n^{2}$ inside the resonator and $1$
outside, $n$ being the index of refraction of the resonator
material. The upper limit for the radial integral, $\rho_0(z)$, is
chosen such that it coincides with the first zero of the radial
wave function outside the resonator, $\Phi_m(\rho_0,z)\equiv 0$,
in order to include the effects of the evanescent field. For the
bottle mode of fig.~\ref{fig:3DIntensity} we find $V_{76,80}=690\
\mu\mathrm{m}^{3}$. This value is only twice as large as the mode
volume of a microsphere with $\varnothing\approx 50\ \mu$m.

\begin{figure}[t]
          \centering
          \includegraphics [scale=0.19]{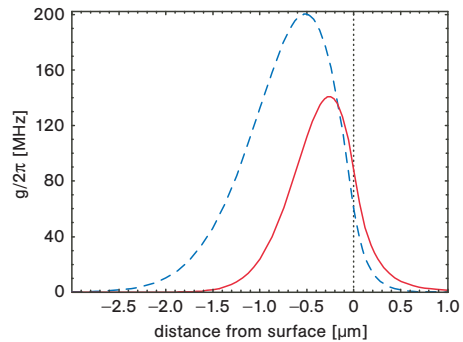}
          \caption{Coupling strength of a Cs atom to the mode of
          the bottle resonator as a function of the distance from
          the resonator surface (solid line). The atom is placed
          at the point of highest field intensity along the the
          resonator axis, $z=z_\mathrm{c}$. For
          comparison, the coupling to an equatorial WGM of a
          $\varnothing = 50\ \mu$m microsphere is also given
          (dashed line). Due to its smaller mode volume, the
          maximum coupling coefficient for the microsphere is a
          factor of 1.4 larger. However, the coupling at
          the resonator surface is 1.5 times larger for the
          bottle resonator, because its evanescent field is more
          pronounced.}
          \label{fig:gs}
\end{figure}

Now, the {\em maximum} coupling strength of a given dipole emitter
to a single photon in the resonator mode is inversely proportional
to the square root of the mode volume, $g\propto 1/\sqrt V$.
However, for WGMs we can only access the field outside of the
resonator. The actual parameter of interest is therefore the
coupling strength at the surface of the resonator. We have
calculated this value for our bottle resonator at the position of
the caustic $z_\mathrm{c}$. For the D2 transition of a Cs atom we
find $g/2\pi \approx\ 90$~MHz. In spite of the larger mode volume,
this exceeds the expected coupling strength to the equatorial WGM
at the surface of a $\varnothing\approx 50\ \mu$m microsphere by a
factor 1.5. The physical reason for this result lies in the fact
that, due to the smaller diameter of the bottle resonator, its
evanescent field is more pronounced than for the larger
microsphere (see fig.~\ref{fig:gs}).

We now examine the mode spectrum of the bottle resonator, given by
eq.~(\ref{eq:eigenwavenumber}). For our geometry, i.e.~$R_0 \Delta
k\ll 1$ and $m\approx q$, one finds that the frequency spacing
between modes for which the azimuthal quantum number $m$ differs
by one approximately equals $\Delta\nu_m\approx c/2\pi n
R_0\approx 4$~THz, just as large as for a microsphere resonator of
the same radius. However, the mode spacing concerning the axial
quantum number $q$ is much smaller and is given by
$\Delta\nu_q\approx c\Delta k/2\pi\approx 150\ $GHz, one order of
magnitude smaller than for a $\varnothing\approx 50\ \mu$m
microsphere. Tuning over one free spectral range therefore only
requires a change of the optical path length of $4\times 10^{-4}$.
This can be realized by varying the resonator temperature by 30~K,
a parameter range that is easily accessible in an experiment.

Mechanical strain could also be used to tune the bottle resonator.
Concerning its mechanical properties, we approximate the resonator
as a cylinder of radius $R_0$. This is well justified in our case
and allows an easy estimation of the effect of the strain on the
mode frequency $\nu_{mq}$. Using the Poisson coefficient and the
elasto-optic coefficients of silica \cite{Borrelli} and neglecting
the modification of the curvature $\Delta k$ along the resonator
axis, we find
\begin{equation}\label{eq:strain_tuning}
    \Delta\nu_{mq}/\nu_{mq}\approx -\Delta R/R_0
    -\Delta n/n\approx-0.2\, \Delta L/L
\end{equation}
for the  TM mode (polarization parallel to the strain), where $L$
is the length of the resonator. One FSR tuning in our case thus
requires a length change of about $2\times 10^{-3}$. Using the
value of $7.2\times 10^{10}$~Pa for the elasticity module of
silica, this implies a strain of about 0.15~GPa, more than one
order of magnitude smaller than the typical damage threshold for
silica fibers of 3~GPa \cite{Glaesmann}. Strain tuning over
several FSR therefore seems possible.

Apart from the advantageous properties concerning mode volume and
tunability, the bottle resonator geometry and the resulting
spatial structure of the modes also offer interesting advantages
with respect to their handling and possible applications. Since
bottle resonators exhibit two spatially well separated regions of
enhanced field strength (see fig.~\ref{fig:3DIntensity}), they are
true two-port devices, contrary to microspheres. One of these two
caustics could therefore be used to couple light into and out of
the resonator by means of a thin unclad optical fiber
\cite{Knight}, while a dipole emitter could be coupled to the
other caustic.

In practice, it has been demonstrated that a prolate resonator
geometry with the dimensions considered here can be realized on
unclad optical fibers using fiber pulling techniques and CO$_2$
laser microstructuring \cite{Karantzas}. In order to obtain the
desired high $Q$ values, these processes will have to be optimized
to yield a surface quality that minimizes scattering losses.
Furthermore, the bulk absorption of the fiber cladding, which
would exclusively guide the light in this case, has to be
sufficiently low. Using low-loss telecommunication fibers, we are
however confident, that this condition can be fulfilled.

Summarizing, we have studied the properties of whispering gallery
modes in highly prolate shaped silica resonators. We have shown
that such ``bottle resonators'' sustain modes that exhibit two
spatially well separated caustics with an enhanced field strength.
For a distance of 140~$\mu$m between the caustics and a resonator
diameter of 16~$\mu$m we find a mode volume of 690~$\mu$m, only
twice as large as for an equatorial whispering gallery mode in a
silica microsphere with a diameter of 50~$\mu$m. In spite of this
larger mode volume, the coupling strength of an atom to the
evanescent field of the mode {\em outside} of the bottle resonator
is larger than for a microsphere: Due to the smaller radius of the
bottle resonator, its evanescent field is stronger. For the D2
transition of Cs, we calculate a coupling strength of $g/2\pi
\approx\ 90$~MHz, much larger than the atomic line width and the
expected cavity line width for an estimated quality factor of
$10^8$--$10^9$. The bottle resonator should therefore allow to
enter the strong coupling regime in neutral atom cavity QED.

At the same time, we have shown that the mode spacing for the
bottle resonator is one order of magnitude smaller than for a
microsphere of 50~$\mu$m diameter. By varying the temperature or
applying mechanical strain to the resonator, tuning over more than
one free spectral range should therefore be possible. Moreover,
the fact that bottle resonators are true two-port devices is an
important advantage when it comes to probing the light field in
the resonator mode while simultaneously coupling a dipole emitter
to it. Finally, it has recently been proposed to trap and guide
atoms around thin unclad optical fibers using a two-color
evanescent light field \cite{Kien}. Our resonator design seems
particularly suited to be combined with such a surface trap in
order to couple trapped cold atoms to the resonator mode.

We acknowledge support from the Deutsche Forschungsgemeinschaft in
the framework of the research unit 557 ``Light Confinement and
Control with Structured Dielectrics and Metals''.

\end{document}